\documentclass[aps,prb,twocolumn,superscriptaddress,showpacs]{revtex4-2}
\usepackage[utf8]{inputenc}
\usepackage{longtable}
\usepackage{amsmath}%
\usepackage{amsfonts}%
\usepackage{amssymb}%
\usepackage{graphicx}
\usepackage[hidelinks, breaklinks=true]{hyperref}
\usepackage{CJKutf8} 
\usepackage{ulem} 
\usepackage[product-units=power, separate-uncertainty=true, multi-part-units=single]{siunitx} 
\usepackage{xspace} 
\usepackage{xcolor}

\newcommand{\myref}[7]{\href{http://dx.doi.org/#7}{#1, #2, #3 \textbf{#4}, #5 (#6).}} 

\newcommand{\CSiC}{C$_\mathrm{SiC}$\xspace}
\newcommand{\SiSiC}{Si$_\mathrm{SiC}$\xspace}
\newcommand{\sqthree}{$\left(\sqrt{3}\times\sqrt{3}\right)R\ang{30}$\xspace}
\newcommand{\sixsqthree}{$\left(6\sqrt{3}\times6\sqrt{3}\right)R\ang{30}$\xspace}
\newcommand{\ZLG}{ZLG\xspace}
\newcommand{\SbOx}{SbO$_\mathrm{x}$\xspace}

\newcommand{\Pc}{$P_\mathrm{c}^\mathbf{H}$\xspace}
\newcommand{\Fc}{$F_\mathrm{c}^\mathbf{H}$\xspace}

\newcommand{\etal}{\textit{et al.}\xspace}

\newcommand{\blue}[1]{#1}
\newcommand{\bsout}[1]{}

\newcommand{\pgi}{Peter Grünberg Institut (PGI-3), Forschungszentrum Jülich, 52425 Jülich, Germany}
\newcommand{\jara}{Jülich Aachen Research Alliance (JARA), Fundamentals of Future Information Technology, 52425 Jülich, Germany}
\newcommand{\rwth}{Experimentalphysik IV A, RWTH Aachen University, 52074 Aachen, Germany}
\newcommand{\TUC}{Institute of Physics,  Faculty of Natural Sciences, TU Chemnitz, 09107 Chemnitz, Germany}
\newcommand{\MAIN}{Center for Materials, Architectures and Integration of Nanomembranes (MAIN), TU Chemnitz, 09107 Chemnitz, Germany}
\newcommand{\diam}{Diamond Light Source Ltd., Didcot, OX110DE, Oxfordshire, United Kingdom}

\begin{document}

\title{Vertical structure of Sb-intercalated quasifreestanding graphene on SiC(0001)}

\author{You-Ron~Lin (\begin{CJK*}{UTF8}{bsmi}林~又~容\end{CJK*})}      \affiliation{\pgi} \affiliation{\jara} \affiliation{\rwth}
\author{Susanne~Wolff} \affiliation{\TUC} \affiliation{\MAIN}
\author{Philip~Schädlich} \affiliation{\TUC} \affiliation{\MAIN}
\author{Mark~Hutter}      \affiliation{\pgi} \affiliation{\jara} \affiliation{\rwth}
\author{Serguei~Soubatch}    \affiliation{\pgi} \affiliation{\jara} 
\author{Tien-Lin~Lee (\begin{CJK*}{UTF8}{bsmi}李~天~麟\end{CJK*})}    \affiliation{\diam} 
\author{F.~Stefan~Tautz}   \affiliation{\pgi} \affiliation{\jara} \affiliation{\rwth}
\author{Thomas~Seyller} \affiliation{\TUC} \affiliation{\MAIN}
\author{Christian~Kumpf}      \affiliation{\pgi} \affiliation{\jara} \affiliation{\rwth}
\author{Fran\c{c}ois~C.~Bocquet} \email{f.bocquet@fz-juelich.de (he/him/his)} \affiliation{\pgi} \affiliation{\jara}  
\date{\today}

\begin{abstract}
Using the normal incidence x-ray standing wave technique as well as low energy electron microscopy we have investigated the structure of quasi-freestanding monolayer graphene (QFMLG) obtained by intercalation of antimony under the \sixsqthree reconstructed graphitized 6H-SiC(0001) surface, also known as zeroth-layer graphene. We found that Sb intercalation decouples the QFMLG \bsout{very} well from the substrate. The distance from the QFMLG to the Sb layer almost equals the expected van der Waals bonding distance of C and Sb. The Sb intercalation layer itself is mono-atomic, \bsout{very} flat, and  located much closer to the substrate, at almost the distance of a covalent Sb-Si bond length. All data is consistent with Sb located on top of the uppermost Si atoms of the SiC bulk. 
\end{abstract}

\maketitle
\section{Introduction}
The groundbreaking work of Novoselov and Geim \etal \cite{Novoselov2005} inspired numerous scientists to investigate the properties and potential applications of the fascinating 2D material graphene. In particular, the concept of epitaxial graphene grown on the surfaces of hexagonal silicon carbide (SiC) polytypes as introduced by Berger \etal \cite{Berger2004} received considerable attention. Since this early work, epitaxial graphene growth has been improved by the introduction of Argon-assisted sublimation growth \cite{Virojanadara2008,Emtsev2009} and recently by the development of the polymer-assisted sublimation growth \cite{Kruskopf2016}. The electronic and structural properties of epitaxial graphene make it a promising material for applications like, for example, high frequency transistors \cite{YMLin2009}, quantum metrology \cite{Tzalenchuk2010} or THz detectors \cite{Bianco2015,Schlecht2019}. 

The growth of graphitic layers on SiC(0001) has been studied experimentally since 1975 \cite{VanBommel1975,Forbeaux1998}. It is known that SiC decomposes at temperatures above $1150^\circ$C, at which the volatile Si atoms sublimate, whereas the carbon atoms arrange in a \sixsqthree superstructure with respect to the SiC substrate. This carbon layer, which we denote as the zeroth-layer graphene (\ZLG), is strongly coupled to the SiC substrate and lacks the typical $\pi$-bands of graphene \cite{Mattausch2007, Emtsev2008} while exhibiting the same lateral structure as monolayer graphene (MLG).
	
When SiC is annealed, \ZLG is formed first. At higher preparation temperature ($> 1250^\circ$C), the ZLG transforms into graphene, as indicated by the appearance of its characteristic electronic band structure, since a new ZLG is formed underneath, which decouples the first one from the substrate \cite{Emtsev2008, Hannon2011}. In this way, one can prepare MLG and even stacked graphene multilayers. However, the \ZLG influences the electronic properties of the graphene layers above, resulting in an n-type doping with a Dirac point energy $E_\text{D}$ of 450~meV \cite{Ohta2007} and an undesirable temperature dependence of the carrier mobility \cite{Speck2011}.

An elegant way to remove the electronic properties induced by the \ZLG is to decouple it from the SiC substrate by intercalation. Thereby, a certain chemical species (intercalant) is brought between the SiC(0001) surface and the \ZLG, which decouples the latter from the substrate and transforms it into so-called quasi-freestanding monolayer graphene (QFMLG). 

A prominent example is the intercalation with hydrogen \cite{Riedl2009,Virojanadara2010}, which produces a decoupled QFMLG on a H-saturated SiC(0001) surface \cite{Sforzini2015}. Such a QFMLG shows little temperature dependency in charge carrier mobility \cite{Speck2011} and an excess of holes due to the spontaneous polarization of the hexagonal SiC substrate \cite{Ristein2012,Mammadov2014,Slawinska2015}. Several different chemical species have been used to intercalate ZLG \cite{Briggs2019a}. Depending on the type of intercalant and on its amount, it is possible to induce interesting electronic properties of QFMLG. For example, ambipolar doping of the graphene layer was observed upon germanium intercalation \cite{Emtsev2011}, where the charge carrier type depends on the amount of intercalated Ge. This effect can also be seen for gold intercalation \cite{Gierz2010}. Another example is lithium- or calcium-doped graphene showing superconductivity \cite{Profeta2012}. For bismuth-intercalated graphene, depending on the amount of the intercalated material, two phases of different crystalline and electronic structures can be distinguished \cite{Sohn2021}.

Recently a new method for intercalation using solid elements with high vapor pressure was reported \cite{Wolff2019}. In this process, intercalation under the \ZLG was achieved by annealing the Sb-covered SiC(0001) sample in Ar at atmospheric pressure \cite{Wolff2019}. X-ray photoelectron spectroscopy (XPS) verified the decoupling of the \ZLG and the formation of a QFMLG \cite{Wolff2019}. Sb~3$d$ core level spectra indicated a layer of metallic Sb at the interface between SiC(0001) and QFMLG \cite{Wolff2019}. The reported changes in the low-energy electron diffraction (LEED) pattern in Ref.~\cite{Wolff2019}, namely a strong reduction of the Moiré spots associated with the ZLG, supported the decoupling observed in XPS. This is typical for ZLG intercalation on SiC and was observed, for instance, for H \cite{Riedl2009}, Au \cite{Gierz2010}, Ge \cite{Emtsev2011}, Si \cite{Xia2012}, Cu \cite{Forti2016}, and Yb \cite{Rosenzweig2019} intercalants. Using angle-resolved photoelectron spectroscopy (ARPES) it was shown that QFMLG formed in this way is n-doped \cite{Wolff2019}. However, no information about the vertical structure of this system has been reported so far, such as the thickness of the Sb layer or its vertical distances to the graphene layer and to the substrate, although such information would be highly relevant, because the layer distance is \bsout{an excellent} \blue{a good} indicator of interaction strength \cite{Silva2019, Silva2018, Weiss2018, Sforzini2016, Sforzini2015, Runte2014}. The thickness of the intercalated layer can affect the electronic properties and the possible application of the system, as in the case of two dimensional Ag layers on epitaxial graphene \cite{Briggs2019a}. In this work, we unambiguously confirm the success and homogeneity of the Sb decoupling using low energy electron microscopy (LEEM) and normal incidence x-ray standing wave (NIXSW). The latter also reveals the vertical structure of QFMLG -- intercalated by metallic Sb -- with a precision better than $\pm 0.2$~Å.

\section{Experimental Methods}
\subsection{Sample preparation}
We used nitrogen-doped 6H-SiC(0001) wafers purchased from SiCrystal GmbH, Germany. Prior to the graphitization process the SiC substrate was etched in hydrogen atmosphere~\cite{Ostler2010}. The \ZLG growth process was then performed in a furnace in Ar atmosphere at a pressure 1000~mbar and an Ar flow rate of 0.1~slm. By annealing the substrate at a temperature of 1475$^\circ$C for 15~min a \ZLG sample was obtained as described in Ref.~\cite{Ostler2010}. Because Sb has a high vapor pressure, a thick layer of Sb (50~nm) must be deposited first on top of the \ZLG sample. This was performed in a separate system by molecular beam epitaxy using a calibrated Sb Knudsen cell \cite{Wolff2019}. During the deposition process, the pressure was maintained at or below $1 \times 10^{-9}\,\text{mbar}$. Since annealing in UHV leads to desorption of Sb before intercalation, it was necessary to perform the intercalation \textsl{ex-situ} by annealing the Sb-covered \ZLG sample in Ar atmosphere at a pressure of 1000~mbar and an Ar flow rate of 0.1~slm \cite{Wolff2019}. The same furnace was used for the for \ZLG growth~\cite{Ostler2010} and Sb intercalation. A consecutive two-step annealing at $400\,^\circ\text{C}$ for $30\,\text{min}$ and at $550\,^\circ\text{C}$ for $60\,\text{min}$ was used to intercalate Sb and obtain QFMLG \cite{Wolff2019}. Note that most of the Sb originally deposited on top of the ZLG desorbs from the surface during this process. Only a small part of the Sb is needed to form a homogeneous intercalation layer at the interface between SiC and QFMLG.

After intercalation, and again after transport of the samples to the synchrotron, LEED and ARPES (He I) measurements were performed to confirm the quality of the QFMLG. In all cases, the obtained data was of comparable quality as those shown in Fig.~4(a) and Fig.~5(b) of Ref.~\cite{Wolff2019}.

\subsection{LEEM}
Low-energy electron microscopy (LEEM) was performed with a SPECS FE-LEEM P90 instrument, on samples transported in air. µ-LEED images are obtained by limiting the incoming electron beam with the help of an aperture to the desired region of interest. From a series of LEEM images recorded as a function of electron energy, LEEM-IV data are extracted for selected regions of the surface \cite{Gopalan2016}.

\subsection{NIXSW}
NIXSW experiments were carried out at the I09 beamline of Diamond Light Source Ltd, Didcot, UK. This beamline is equipped with a Scienta EW4000 HAXPES hemispherical electron analyzer. All XPS and NIXSW data presented in this work were measured at room temperature in normal incidence geometry. Photoelectrons with an emission angle between $60^\circ$ and $90^\circ$ with respect to the surface normal were collected by the electron analyzer. 

For NIXSW \cite{Zegenhagen2013, Zegenhagen1993, Woodruff1998, Woodruff2005}, the x-ray beam energy $h\nu$ is tuned in order to fulfill the Bragg condition for a specific $\mathbf{H}=(hkl)$ (or $(hkil)$ for a hexagonal lattice) reflection of the crystalline sample substrate, the 6H-SiC(0006) in our case. A standing wave field is formed under these conditions by the interference of the incoming and the back-diffracted beams. The period of the standing wave is defined by the bulk lattice spacing $d_{hkl}$ ($2.52$~\AA\ in our case), and its phase $\Phi(h\nu)$ varies with the photon energy around the Bragg condition. In an NIXSW measurement, $h\nu$ is scanned through the Bragg condition (in our case within a $\pm 2$~eV interval around the Bragg energy), changing the phase $\Phi(h\nu)$ from $\pi$ to $0$, and hence shifting the nodes and anti-nodes of the standing wave by $\frac{1}{2} d_{hkl}$ inward along the $(hkl)$ direction. At each photon energy step of such a scan, photoelectron spectra of all relevant atomic species are recorded. The photoelectron yield curves $Y(h\nu)$ are then obtained by plotting the integrated intensities from the individual spectra vs.\ photon energy. Each yield curve shows a characteristic modulation that is related to the vertical position of the corresponding atomic species, and is fitted by 
\begin{eqnarray}
	Y(h\nu) &=& 1 + S_R R(h\nu) \label{eq:Yield}  \\
	&+& 2|S_I|\sqrt{R(h\nu)} F_\mathrm{c}^\mathbf{H} \cos(\Phi(h\nu)-2\pi P_\mathrm{c}^\mathbf{H} + \psi). \nonumber
\end{eqnarray}
Here, $R(h\nu)$ is the x-ray reflectivity, and $S_I=|S_I| \mathrm{e}^{-i\psi}$ and $S_R$ are parameters for the correction of non-dipolar effects and a deviation from perfect normal incidence that arises from the experimental conditions \cite{Bocquet2019, vanStraaten2018}. 
The fit parameters that are obtained from Eq.~\ref{eq:Yield} are the coherent position \Pc and the coherent fraction \Fc, both ranging from 0 to 1. \Fc is a measure of the vertical order. \Fc$ = 1$ indicates that all atoms of the considered species are located in a well-ordered manner at the same distance relative to a Bragg plane (although not necessarily relative to the same Bragg plane). \Pc represents the position of the probed atomic species with respect to the next Bragg plane below, in units of $d_{hkl}$. In case of adsorbates on a surface, the actual adsorption height can be calculated from the coherent position by $z = d_{hkl}(n+$\Pc), where $n$ is the number of Bragg planes being located between the surface and the atomic species considered. 

For all distinguishable species, the photoelectron yield curves were extracted from the individual XPS spectra using \textsc{CasaXPS} \cite{CasaXPS}. The obtained yield and reflectivity curves were further analyzed using the dedicated NIXSW analysis software \textsc{Torricelli} \cite{Bocquet2019, Torricelli}. For more details on the data analysis, in particular the error analysis, the correction of non-dipolar effects, and the deviation from perfect normal incidence, see Refs. \cite{Mercurio2013, Bocquet2019, vanStraaten2018}. 

\section{Results and Discussion}
\begin{figure}[t!]
	\includegraphics[width=\linewidth]{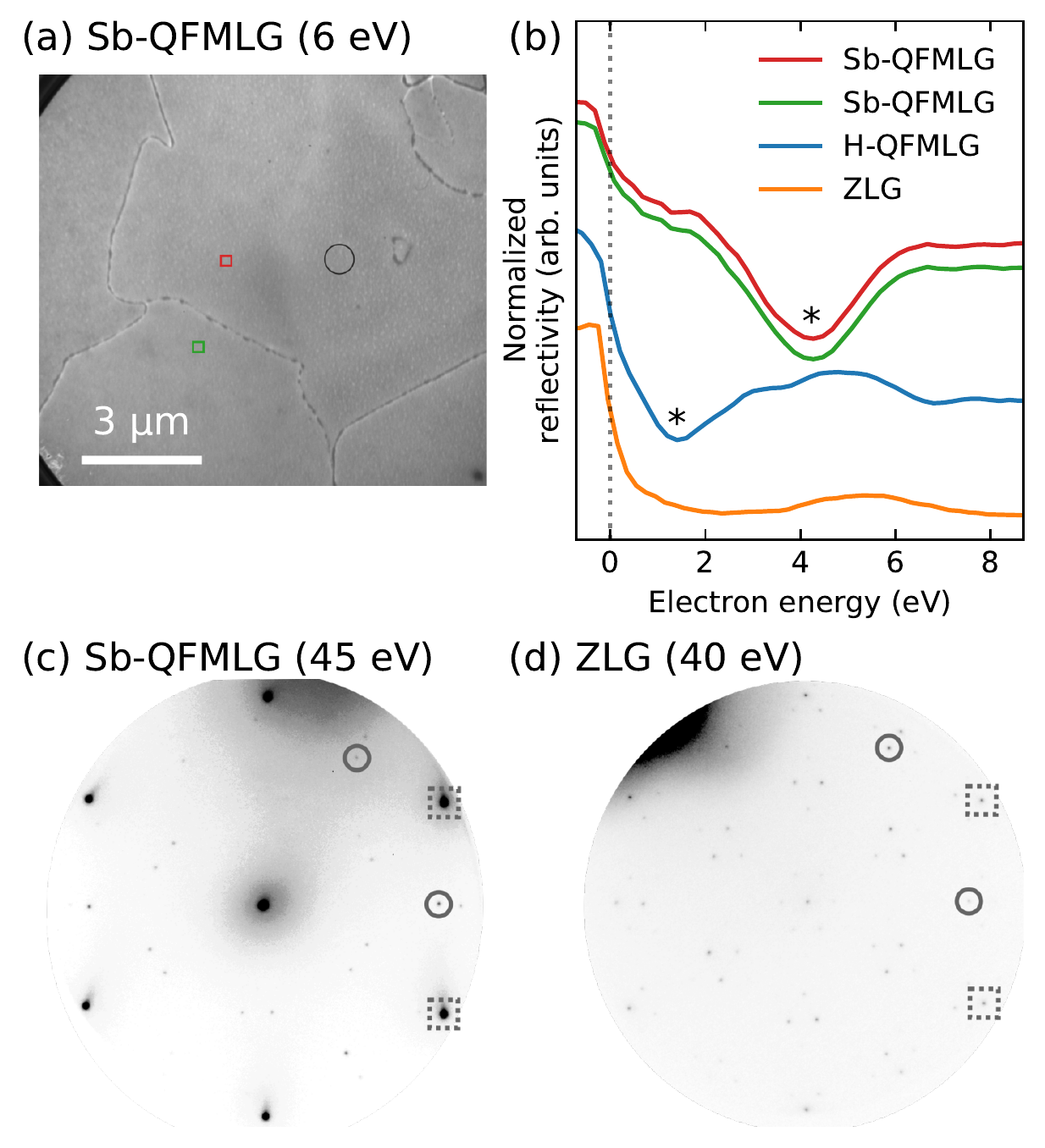}\\
	\caption{\label{fig:LEEM} \blue{(a) LEEM image recorded on Sb-QFMLG. (b) Normalized LEEM-IV spectra (red and green lines) recorded from the areas indicated by squares in (a) with respective colors. For comparison, we show data for H-QFMLG (blue) and ZLG (orange). For the sake of clarity, the curves are vertically displaced. $\mu$-LEED pattern recorded on (c) Sb-QFMLG and (d) ZLG. The Sb-QFMLG $\mu$-LEED pattern was taken from the area indicated by a black circle in (a). We indicated the (01) and (10) diffraction spots of SiC(0001) (gray circles) and graphene (gray dotted squares). In all cases the substrate is 6H-SiC(0001).} }
\end{figure}

In Fig.~\ref{fig:LEEM}(a), a LEEM image of the surface of ZLG after Sb intercalation \blue{(Sb-QFMLG)} is shown. Very large (several $\mu$m$^2$) and homogeneous terraces are visible. In the corresponding $\mu$-LEED pattern (Fig.~\ref{fig:LEEM}(c)), the graphene spots are dominant at this energy, while the SiC spot intensities are very weak. Despite the enhanced gray scale, Moiré spots associated with the ZLG \blue{(see Fig.~\ref{fig:LEEM}(d))} are barely visible \blue{and no additional spot appeared after intercalation}, in agreement with Ref.~\cite{Wolff2019}. In Fig.~\ref{fig:LEEM}(b), LEEM-IV data are compared for Sb-QFMLG, hydrogen-intercalated quasi-freestanding graphene (H-QFMLG) and ZLG. \bsout{, and expitaxial monolayer graphene (EMLG).} They all show a clear minimum in the IV curve \blue{(see the $*$ symbols in Fig.~\ref{fig:LEEM}(b))}. In contrast, the IV curve of the ZLG is rather flat for energies above 1~eV. The appearance of a distinct minimum after intercalation is a well-known signature of the decoupling of the ZLG by intercalation and the formation of monolayer graphene, \blue{see Ref.~}\cite{Hibino2008, Riedl2009, Emtsev2011, Ostler2014} \blue{and references therein}. The shape of the Sb-intercalated QFMLG LEEM-IV curve is therefore a clear indication for a successful intercalation. Note that the precise energy position of the minimum depends on the distance as well as on the effective potential between the graphene layer and the substrate \cite{Hibino2008, Feenstra2013}. In order to reveal whether or not the Sb-intercalated graphene layer is quasi-freestanding, NIXSW is the experimental method of choice.

\begin{figure}[t!]
	\includegraphics[width=\linewidth]{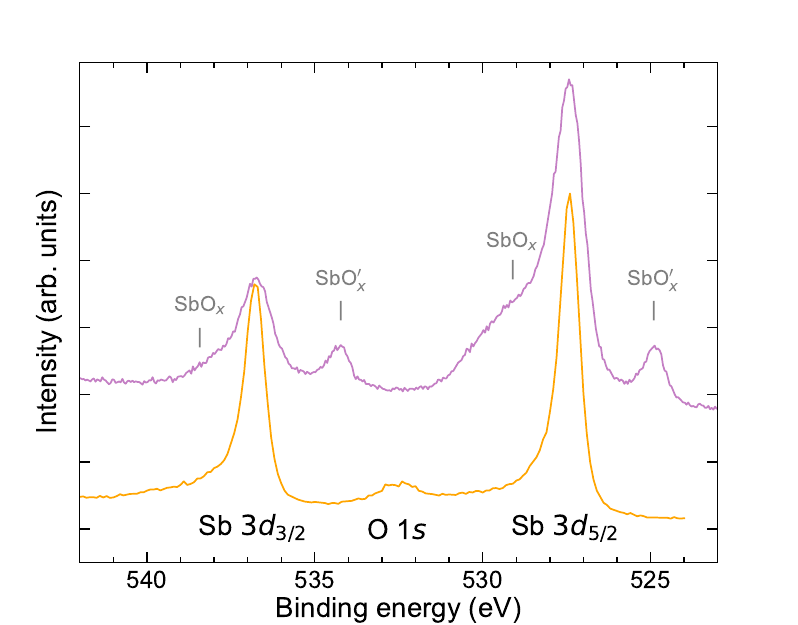}\\
	\caption{\label{fig:XPS} Sb~3$d$ core-level spectra. Orange line: Sb-intercalated quasi-freestanding graphene on SiC(0001) (photon energy of 2458~eV). Magenta line: ZLG sample with incomplete Sb intercalation (photon energy of 700~eV). Two oxidized Sb species are marked.}
\end{figure}

We have recorded core-level spectra and NIXSW data for all relevant species (Sb, C and Si) for Sb-intercalated quasi-freestanding graphene. In the data analysis, an important first step is the identification of all relevant components in the spectra. In the Sb~3$d$ data, beside the 3$d_{5/2}$ and 3$d_{3/2}$ doublet (peaks at $527.8$~eV and $536.5$~eV, respectively, see the orange curve in Fig.~\ref{fig:XPS}), we find an additional small signal at $532.5$~eV that we attribute to oxygen. One might suspect that this little amount of oxygen is located in the Sb intercalation layer, thus forming \SbOx. However, in this case a doublet of (chemically shifted) Sb 3d peaks should be present \cite{Wolff2019}, as it is the case for in the magenta spectrum in Fig.~\ref{fig:XPS}, recorded on an incompletely intercalated sample after air exposure. In the orange spectrum, such a chemically shifted doublet is not present, which lets us conclude that no significant amount of \SbOx is formed on completely Sb-intercalated quasi-freestanding graphene. Therefore, the oxygen peak in our spectrum most likely originates from residual oxygen atoms at step edges and step bunches where the ZLG or QFMLG do not perfectly protect the SiC substrate from oxidation. Any conclusion drawn in this paper for the homogeneous Sb-intercalated regions of the surface are thus not affected by the residual oxygen.

\begin{figure}[t]
	\includegraphics[width=\linewidth]{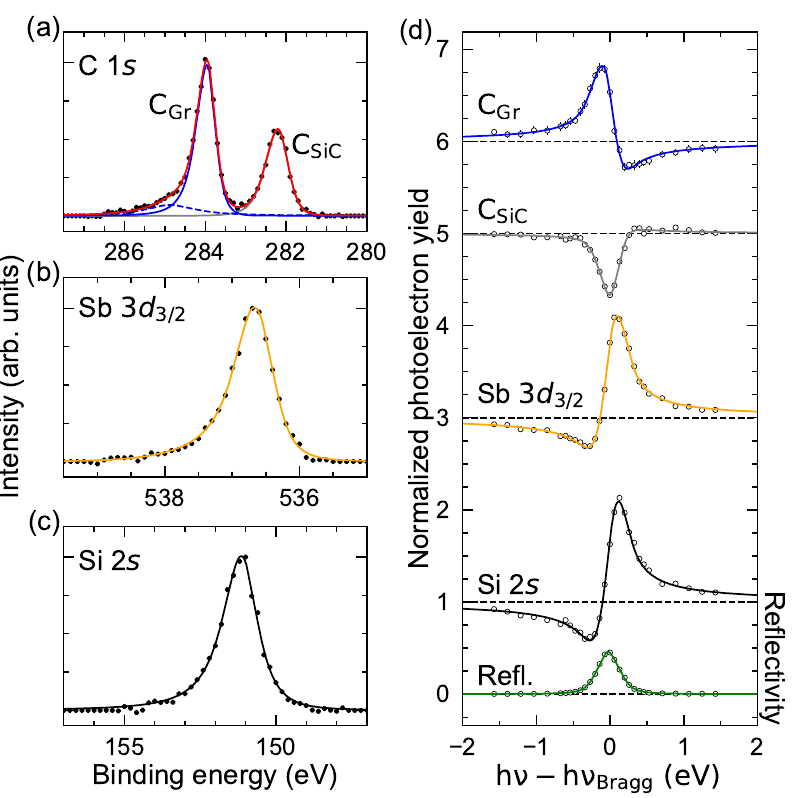}\\
	\caption{\label{fig:yieldcurves} (a)-(c) C~1$s$, Sb~3$d_{3/2}$, and Si~2$s$ core-level spectra recorded on Sb-intercalated quasi-freestanding graphene on SiC(0001)  at a photon energy of $2465$~eV, that is $4$~eV below the SiC(0006) Bragg energy. A linear background was subtracted. (d) Corresponding NIXSW yield curves and a typical reflectivity of the SiC(0006) reflection. The curves for C$_\mathrm{Gr}$, \CSiC and Sb~3$d_{3/2}$ are vertically displaced by $5$, $4$ and $2$, respectively.}
\end{figure}

For the NIXSW experiments, we used only the Sb~3$d_{3/2}$, since the O~1$s$ peak is sufficiently far away on the lower binding energy side and the ratio between the 3$d_{3/2}$ and 3$d_{5/2}$ peaks is fixed to 2/3 (see Table 3.2 in Ref.~\cite{Hofmann2013}). Typical spectra for all core-levels (recorded using hard x-rays) are shown in Fig.~\ref{fig:yieldcurves}(a-c). In the C~1$s$ spectrum we see two well-separated peaks stemming from QFMLG and bulk SiC. We used asymmetric Lorentzian functions to fit all core-levels after subtraction of a linear background. The full width at half maximum (FWHM) values of the best fits are: 0.46~eV for the C$_\mathrm{Gr}$ component, 0.56~eV for C$_\mathrm{SiC}$, 0.62~eV for Sb~3$d_{3/2}$ and 1.21~eV for Si~2$s$.
The resulting photoelectron yield curves are shown in Fig.~\ref{fig:yieldcurves}(d), including fits performed using the software package \textsc{Torricelli} \cite{Bocquet2019, Torricelli}, considering both non-dipolar effects in the photoemission process and a small (unavoidable) deviation from normal incidence ($\theta = 86.5^\circ$ in our case) \cite{vanStraaten2018}. Note that a proper correction of non-dipolar effects can only be performed for photoemission from an $s$-state \cite{Vartanyants2013, Woodruff2005}. Hence, for Sb~$3d$ the data are uncorrected, but the correct polarization factor (differing from unity due to $\theta \neq 90^\circ$) was taken into account. For more details on these corrections and on the treatment of the experimental uncertainties see Ref.~\cite{vanStraaten2018,Bocquet2019}.

\begin{table}[b]
  \caption{\label{table:XSWresults} Average results of all individual NIXSW scans on Sb-intercalated quasi-freestanding graphene on SiC(0001). Parameters used in the fitting: Bragg angle $\theta = 86.5^\circ$, electron emission angle relative to the incident synchrotron beam $\phi = 80.9^\circ$, non-dipolar correction factors $\gamma_{C1s} = 1.022$, $\Delta_{C1s} = -0.218$, $\gamma_{Si2s} = 0.707$, and $\Delta_{Si2s} = 2.645$, for details see \cite{Torricelli, vanStraaten2018}. $n$ is the number of lattice planes to the surface.}
  \begin{tabular}{lrrrr}
    \hline \hline
    &  $P_\mathrm{c}$~~~  &  $F_\mathrm{c}$~~~   & $n$ & $z$ (\AA)~ \\ 
    \hline 
    C$_\mathrm{Gr}~1s$   & $0.47(1)$ & ~$0.71(7)$ & $~~2$ & $6.19(2)$  \\
    Sb~3$d_{3/2}$~ ~     & $0.03(1)$ & ~$1.09(2)$ & $~~1$ & $2.58(2)$  \\
    Si$_\mathrm{SiC}~2s$ & $0.01(1)$ & ~$1.12(2)$ & $~~0$ & $0.00~~$  \\
    C$_\mathrm{SiC}~1s$  & $0.76(1)$ & ~$1.02(2)$ &  $-1$ & $-0.63(2)$ \\
    \hline \hline
  \end{tabular}
\end{table}

Note that several such yield curves were measured for C~1$s$, Sb~3$d_{3/2}$, and Si~2$s$ core levels on different positions on the sample surface. In this manner, the homogeneity of the sample was verified by XPS. The averaged results from these multiple scans are listed in Table~\ref{table:XSWresults}. It is remarkable that the intercalated Sb layer shows a very high coherent fraction of $1.09(2)$, comparable with the values obtained for the bulk species \CSiC and \SiSiC. We assign the nonphysically high value (above 1.0) to non-dipolar effects that cannot be corrected for $d$-shell emission \cite{vanStraaten2018}, and to a known non-linearity of the electron analyzer for high count rates (still below the saturation threshold) \cite{Reber2014}. In either case, coherent positions are not affected, and regardless of being slightly larger than 1, the high coherent fraction of the Sb~3$d_{3/2}$ yield does indicate a high vertical order of the Sb atoms. In other words, a very flat Sb layer is formed by the intercalation process. Our data thus excludes the formation of a Sb multilayer. This is in agreement with the fact that we do not see any shifted component in the Sb~$3d$ spectra, which would indicate Sb being located in different chemical environments.

The graphene layer exhibits a significantly smaller coherent fraction of $0.71(7)$. It should be mentioned that for hydrogen intercalated graphene a very similar value of \Fc$ = 0.68$ has been found \cite{Sforzini2015}. We therefore interpret the coherent position obtained for the C$_\mathrm{Gr}$ atomic species as the adsorption height of the graphene layer above the Sb intercalation layer. In principle, the reduction of the coherent fraction of graphene below 1 could be due to a slightly incomplete intercalation, which would cause domains of graphene being located at different heights. But since we did not have any other indication for an incomplete intercalation (see discussion on SbO$_\mathrm{x}$ above), we conclude that an appreciable buckling of the graphene layer contributes to the reduced coherent fraction.

\begin{figure}[t!]
	\includegraphics[width=\linewidth]{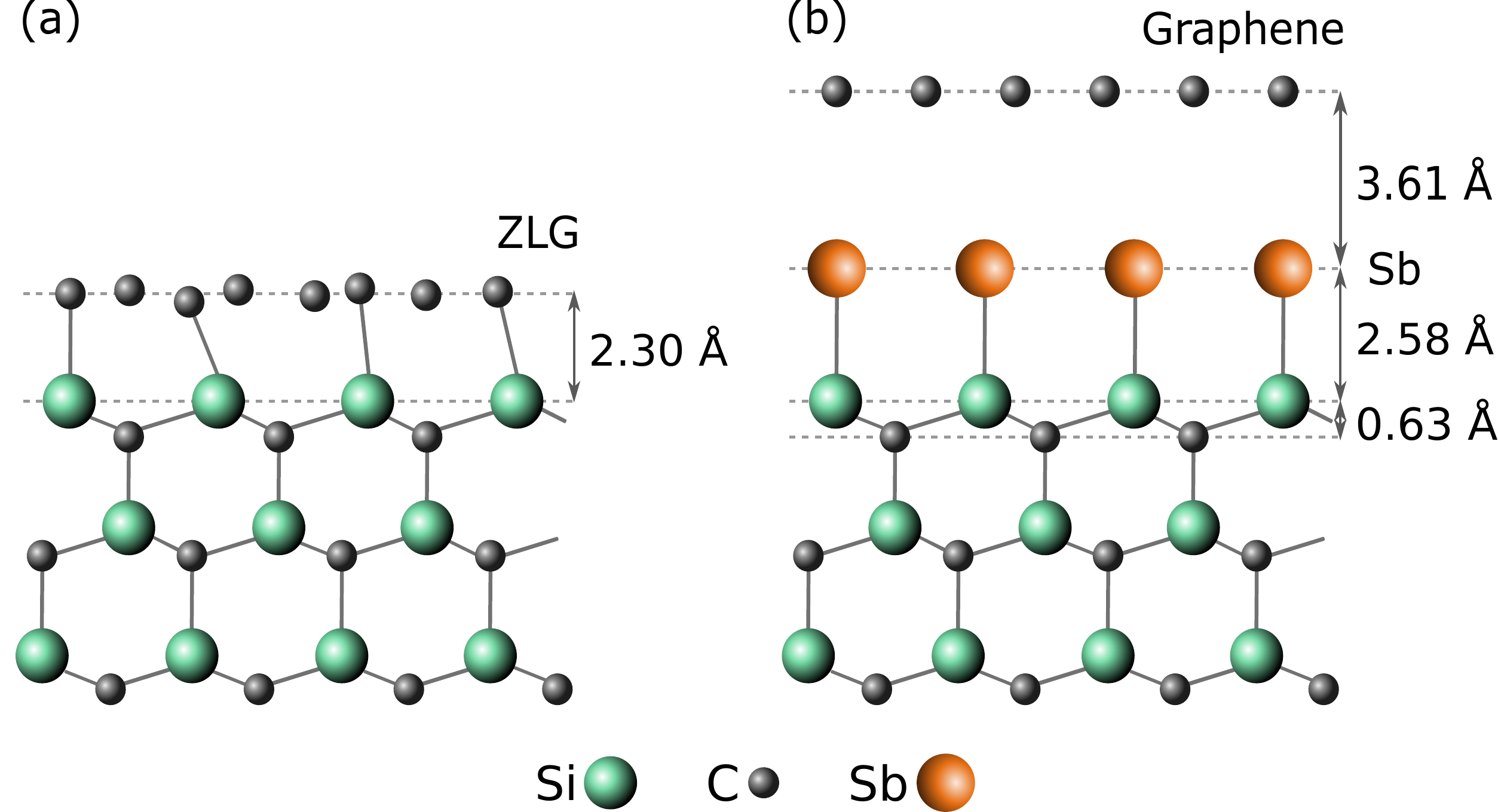}\\
	\caption{\label{fig:structure} Ball-and-stick models of the vertical structure of (a) ZLG before intercalation, from Ref.~\cite{Sforzini2016}, and (b) of Sb-intercalated quasi-freestanding graphene on SiC(0001).}
\end{figure}


A structure model based on our NIXSW data is displayed in Fig.\ \ref{fig:structure}(b). It shows the intercalation layer at a height of $z_{\mathrm{Sb}} = 2.58(2)$~\AA\ above the topmost Si atomic plane, in other words, at a distance very close to the sum of the covalent radii of Si ($1.16$~\AA) and Sb ($1.40$~\AA) \cite{Pyykko2008} and in good agreement with the Sb-Si bond length observed in tetrakis(trimethylsilyl)distibane ($2.594$~\AA) \cite{Becker1982}. This measured distance corresponds to only $\approx 62$\% of the typical van der Waals bonding distance between these two species ($4.16$~\AA) \cite{Mantina2009}, and clearly indicates the chemical interaction between the Sb layer and the SiC substrate. It also suggests that all Sb atoms are located at on-top sites of the uppermost Si atoms (no Sb atoms at other sites), as indicated in Fig.\ \ref{fig:structure}(b). Given the covalent Si-Sb bonding distance, a position on a 3-fold hollow site of the SiC substrate would result in a smaller height for Sb of $\approx 2.28$~\AA. 

Our tentative conclusion that all Sb atoms are occupying on-top sites of the uppermost Si atoms of the SiC(0001) surface, saturating their dangling bonds, is consistent with the fact that LEED shows no indications of any surface reconstruction, but the ($1\times1$) surface unit mesh of the Sb-terminated SiC surface (only diffraction spots of graphene and the substrate are visible [28]).  It is furthermore supported by DFT calculations by Hsu \etal \cite{Hsu2013}, showing the Sb intercalant to be stable at on-top sites of the SiC(0001) substrate. 
A similar scenario, Sb saturating the dangling bonds of the top Si atoms, is found for a monolayer Sb on Si(111). Although the surface is \sqthree reconstructed in this case, the bonding distance between Sb and Si ($2.74$~\AA~\cite{Nakatani1992}) is similar to what we observe here.

The distance between the graphene layer and the Sb intercalation-layer is much larger, $3.62(4)$~\AA. This value corresponds to about $96$\% of the van der Waals bonding distance ($3.76$~\AA\ \cite{Mantina2009}) and is clearly higher than the expected covalent bonding distance ($2.16$~\AA\ \cite{Pyykko2008}).  In very good agreement with our observation, DFT calculations by Hsu \etal \cite{Hsu2013} predicted a Sb-QFMLG distance of 3.659~{\AA}.
Hence, we conclude that the Sb intercalation layer decouples the QFMLG at least structurally \bsout{very} well from the substrate. The observation of a sharp and well visible Dirac cone \cite{Wolff2019}, indicative of an electronic decoupling, supports this finding of a mere van der Waals-like interaction between the atomic species of the QFMLG and the Sb intercalation layer.

\section{Conclusion}
Using a combination of LEEM, XPS and NIXSW we have investigated quasi-freestanding monolayer graphene obtained by intercalating a zeroth-layer graphene on 6H-SiC(0001) with antimony. Homogeneity and intercalation quality were checked by LEEM and XPS. The vertical structure we found by NIXSW consists of a \bsout{very} flat single layer of Sb, covalently bound to the top Si atoms of the substrate. Multiple Sb layers can clearly be excluded. The measured layer distance of $2.58$~\AA\ and the ($1 \times 1$) LEED pattern suggest an on-top position of Sb above the topmost Si atoms. 

The QFMLG layer above the Sb intercalation layer is located much further away, $3.62$~\AA\ above Sb. This indicates van der Waals bonding only between these layers, and hence a very good decoupling of the QFMLG. The structural parameters obtained from our study are in very good agreement with the theoretical prediction by Hsu \etal \cite{Hsu2013}. 

Data shown in the main text are available at the Jülich DATA public repository \cite{JuelichDAta}.

\begin{acknowledgments}
Y.-R.~L., M.~H., F.~S.~T., C.~K. and F.~C.~B. acknowledge funding by the DFG through the SFB 1083 Structure and Dynamics of Internal Interfaces (project A12). S.~W., P.~S., and T.~S.~acknowledge funding by the DFG through the research unit FOR 5242 (project E1). We thank Diamond Light Source for the access to the I09 beamline (proposal NT18398) that contributed to the results presented here, and the I09 beam-line staff (P. K. Thakur, D. Duncan, and D. McCue) for their support during the experiment.
\end{acknowledgments}

\section*{Author contributions}
T.~S. and F.~C.~B. conceived the research.
S.~W. intercalated the sample and carried out LEED and ARPES measurements at TU Chemnitz.  P.~S. acquired and analyzed the LEEM data.
Y.-R.~L., M.~H., S.~S., T.-L.~L., C.~K., and F.~C.~B. performed the NIXSW experiments, and Y.-R.~L. analyzed the data and made the figures. 
All authors discussed the results, and Y.-R.~L., S.~W., C.~K., and F.~C.~B. wrote the paper, with significant input from F.~S.~T. and T.~S.

\end{document}